\documentclass[12pt]{article}

\usepackage[dvips]{color}
\usepackage{epsfig}
\usepackage{subfigure}
\def\lsim{\mathrel{\rlap{
\lower4pt\hbox{\hskip-3pt$\sim$}}
    \raise1pt\hbox{$<$}}}     
\def\gsim{\mathrel{\rlap{
\lower4pt\hbox{\hskip-3pt$\sim$}}
    \raise1pt\hbox{$>$}}}     

\newcommand{\bc}{\begin{center}}
\newcommand{\ec}{\end{center}}
\newcommand{\be}{\begin{equation}}
\newcommand{\ee}{\end{equation}}
\newcommand{\bea}{\begin{eqnarray}}
\newcommand{\eea}{\end{eqnarray} }
\newcommand{\ba}{\begin{array}}
\newcommand{\ea}{\end{array}}

\newcommand{\vvv}[1]{\vec{#1}}
\newcommand{\vp}{{\bar p}}

\newcommand{\ci}[1]{\cite{#1}}

\newcommand{\re}[1]{(\ref{#1})}

\def\Journal#1#2#3#4{{#1}{\bf #2}, #3 (#4)}

\def\NPB{{\em Nucl. Phys. } \bf B}

\def\JMPE{{\em Int. J. Mod. Phys. } \bf E}

\begin{document}
\begin{center}
{\bfseries QUARK-GLUON EVOLUTION IN EARLY STAGE OF ULTRA-RELATIVISTIC HEAVY-ION COLLISIONS}

\vskip 5mm

V.V.~Skokov$^{1 \dag}$,
S.A.~Smolyansky$^{2}$,  and V.D.~Toneev$^{1}$

\vskip 5mm

{\small
(1) {\it
Bogoliubov Laboratory of Theoretical Physics Joint Institute for Nuclear Research, \\ Dubna, Russia
}
\\
(2) {\it
Physics Department, Saratov State University, Saratov, Russia
}
\\
$\dag$ {\it
E-mail: vvskokov@thsun1.jinr.ru
}}
\end{center}

\vskip 5mm

\begin{center}
\begin{minipage}{150mm}
\centerline{\bf Abstract}
A set of coupled kinetic equations describing in
the Abelian approximation  a mixture of quarks and
self-interacting gluons is formulated and solved numerically. The
model includes the Schwinger-like mechanism for particle creation
in a strong field as well as  two-particle elastic collisions
between all mixture components in the Landau approximation of
small-angle scattering. The process of equilibration at the
initial energy density exhibits a dominant quark creation in the
very early time of interaction. It is shown
that damping of energy density oscillations due to elastic scattering of perturbative
quarks and gluons  is not strong enough
to reach thermodynamic equilibrium in  a reasonable relaxation  time.
A possible account for a such behavior is discussed.
\\[3mm]
\end{minipage}
\end{center}

\vskip 10mm

\section{Effective Lagrangian}

 Quark-gluon plasma (QGP) evolution is considered in a following physics scenario. In early stage of
heavy-ion collisions at ultra-relativistic energy a chromo-electric flux tube may be
stretched between nuclear residuals and this strong field results in a spontaneous vacuum pair
creation (Schwinger mechanism). In its turn, the created particles can influence on this
background field (back reaction process) and also suffer subsequent rescatterings relaxing
 to some equilibrium state, in general. This picture  is treated in
the Abelian approximation based on the kinetic equations (KE) derived in~\ci{1,2}. In our paper
we focus on the relative role of components of quark-gluon mixture in  the pre-equilibrium
stage as well as on the question how fast the relaxation process is.

To describe QGP we  use the following effective model  Lagrangian~\ci{Vinni}:
\bea
\label{Lagrangian}
 L(x)& = &\partial _\mu  \Phi ^* (x)\partial ^\mu \Phi
(x) - m_+^2 \left| {\Phi (x)} \right|^2  - e_+^2 \left| {\Phi (x)}
\right|^4 \nonumber\\
&&+ ie_+ \left\{ {[\nabla_n \Phi ^* (x)]\Phi ^2 (x) - [\Phi ^*
(x)]^2 \nabla_n
\Phi (x)} \right\}  \nonumber\\
&&+ \frac{i} {2}\overline \psi  (x)\gamma ^\mu
\stackrel{\leftrightarrow}{\partial} _\mu \psi (x) - m_- \overline
\psi  (x)\psi (x) \nonumber\\
&& +\frac{e_-}{2}\left( \Phi(x)+\Phi^*(x) \right) \overline \psi
(x)(\gamma n)\psi(x)\,\,,
\eea
where indexes    \ '$+$' \  and \ '$-$'   \ are used for bosons (gluons) and
fermions (quarks),  respectively, and $\nabla_n = n^{\mu}\partial_{\mu}$.

We suppose that the total bosonic field  can be decomposed into
a mean-field contribution $\Phi_0(t)$ and its fluctuations
\bea \label{2.2}
 \Phi(x) = \Phi_0(t) + \varphi(x).
\eea
We consider $\Phi_0$ as a neutral  space homogeneous background field. The field of fluctuations $\varphi(x)$
is complex and corresponds to charged field with vanishing mean value.
The $\phi^4$ term  in our Lagrangian \re{Lagrangian} simulates the gluon self-interaction.
Keeping only the second order terms in
 fluctuations, the corresponding equations of motion read
\bea %
\label{2.6}%
 \left[i\gamma^\mu\partial_\mu-e_-\gamma^\mu n_\mu
\Phi_0(t)-m_-\right]\psi(x)&=&0\,,\\ %
\left[D^*_\mu D^\mu +
m_{+}^2 \right] \varphi(x)&=&0\label{2.7} \,,\\%
\label{2.8}
\partial _0^2 \Phi _0(t)  + m_+^2 \Phi _0(t)  + 4 e_+^2 \Phi _0^3(t)- j_-(t)  -
j_+(t) &=&0\,,
\eea
where the covariant derivative $D_{\mu} = \partial_{\mu} + ie_+ n_{\mu}
\Phi_0(t)$.

Alongside with the Dirac equation \re{2.6} and the Klein-Gordon equation \re{2.7}
we get eq.\re{2.8} for evolution of the background mean field. One sees that
equations of motion are self-con\-sis\-tently coupled: the created particles generate
the currents $j_{\pm} (t)$ which form the background field according to \re{2.8}.
In the mean-field approximation  these currents are given as
\bea%
\label{2.9}%
j_- (t) &=& - e_- < \overline \psi (x)(\gamma n)\psi (x)
>,\\\label{2.10} j_+ (t) &=& - e_+  < i\varphi ^*
(x)\nabla_n \varphi (x) - i(\nabla_n \varphi ^* (x))\varphi (x)\nonumber\\
&& - 2e_+ \Phi _0 (t)\varphi ^* (x)\varphi (x)  >\,.%
\eea%

\section{Kinetic equations}

Starting from these equations of motion one can introduce quasi-particles
and obtain the KE for the single particle distribution function on
the basis of diagonalization of Hamiltonian by the
time-dependent Bogoliubov transformation. This result is exact
in the mean-field approximation for a space homogeneous field.
 In our model, the vector $n_{\mu}$  corresponds
 to Flux Tube Model (FTM) geometry and is chosen to be $(0,0,0,1)$.
 So, we arrive at the following KE:
\bea \label{2.11} \frac{{\partial f_\pm (\vp ,t)}}
{{\partial t}} + e_\pm\sigma(t)\frac{{\partial f_\pm (\vp ,t)}}
{{\partial p_\parallel }} = S_\pm (\vp,t) + C_\pm(\vp,t)\,.
\eea
Here, the  chromo-electric field strength
is $\sigma(t) = - d\Phi_0/dt$, $C_\pm(\vp,t)$ is the collision
integral  and $S_\pm (\vp,t)$ is the source term:
  \bea %
  \label{2.12} %
  S_\pm(\vp
,t) &=& \frac{1}{2} W_\pm(\vp ,t) \int\limits_{- \infty }^t {dt'}
W_\pm(\vp,t,t')\times \nonumber \\
&& \left[1 \pm 2f_\pm(P_\pm(t,t'),t' )\right]\cos \left\{ {2
\int\limits_{t' }^t {d \tau} \omega_\pm(\vp;t,\tau )}
\right\}
\eea
with the  transition amplitude
\bea \label{2.13}
W_\pm (\vp ;t,t' ) = \frac{{e_\pm
\sigma(t' )P_\pm (t,t' )}} {{\omega _\pm^2 (\vp ;t,t')}} \left[
{\frac{\varepsilon_{\pm\perp} } {{P_\pm (t,t')}}}
\right]^{2s_\pm}\,,
\eea
where $s_+ = 0$ for gluons and  $s_- = 1/2$ for
quarks. The quasi-particle energy entering this equation is
\bea \label{2.14} \omega^2_\pm
(\vp,t,t')& =& \varepsilon_{\pm \perp }^2 + P_\pm^2(t,t')\,,
 \eea
 where
$
P_\pm (t,t') = p_\parallel  - e_\pm \int\limits_{t' }^t {d\tau
\sigma(\tau )}
$
and 
$
\varepsilon^2_{\pm \perp} = m_\pm^2  + p^2_\bot.
$
General  features of this source term are described in~\ci{2,3}.

Kinetic equation \re{2.11} may be rewritten in the form of a system of
ordinary differential equations~\ci{3,p}:
\bea \frac{{d f_\pm (\vp ,t)}}
{{d t}}& =& \frac{1}{2}W_\pm(\vp,t) v_\pm(\vec{p},t) + C_\pm(\vec{p},t),
\label{2.19}\\
\frac{{d v_\pm (\vp ,t)}} {{d t}}& =& W_\pm(\vp,t) \ \Big( 1 \pm
2f_\pm(\vec{p},t) \Big) - 2\omega_\pm(\vp) \ u_\pm(\vec{p},t),
\label{2.20}\\
\frac{{d u_\pm (\vp ,t)}} {{d t}}& =& 2{\omega}_\pm(\vp) \
v_\pm(\vec{p},t), \label{2.21}
 \eea
 where two new functions have been introduced :
 \bea%
  \label{2.17}%
   u_\pm(\vec{p},t) &=& \int_0^t dt^{\prime}
W_\pm(\vp,t,t')
\Big( 1 \pm 2f_\pm(P_\pm(t,t'),t') \Big)  \sin[2\int_{t'}^t d\tau \omega_\pm(t,\tau)],
\\ \label{2.18} v_\pm(\vec{p},t) &=& \int_0^t
dt^{\prime} W_\pm(\vp,t,t')
\Big( 1 \pm 2f_\pm(P_\pm(t,t'),t^{\prime}) \Big)
 \cos[2\int_{t'}^t d\tau \omega_\pm(t,\tau)]. %
\eea

\section{Collision integral}
 Neglecting particle-particle collisions does not result in complete picture
 of a collision at large time moments when
the interaction force between particles is getting greater than the mean-field interaction.
However, the direct evaluation of the  CI $ C_\pm(\vec{p},t)$ gives rise to a huge numerical problem.
In earlier papers the CI was introduced in the relaxation time approximation with
 time- and momentum-independent relaxation time what does not allow to restore true
dynamics of the relaxation process and to estimate properly the relaxation time~\ci{2,3}.
Here, the CI will be obtained on a dynamical basis but in a simplified manner by
making use of the Landau approximation, i.e.   assuming small momentum transfer in elastic
$qq$-, $qg$- and $gg$-collisions. The appropriate cross sections are calculated in the
perturbative approximation~\ci{10}:
\bea%
 \label{3.1}%
 \frac{d \sigma_{qq}}{d t} &=& \frac{\pi}{2} \frac{N_c^2 - 1}{\,
 N_c^2}\frac{\alpha_s^2}{ s^2}\left[\frac{s^2+ u^2}{t^2} + \frac{s^2+ t^2}{u^2} -\frac{2\, s^2}{3\, t
 u}\right],  \\
 \label{3.2}%
\frac{d \sigma_{gg}}{d t} &=& 4\pi  \frac{N_c^2}{
  N_c^2-1}\frac{\alpha_s^2}{ s^2}\left[3 - \frac{u t}{s^2} - \frac{u s }{t^2} -\frac{s t}{u^2}\right],  \\
  \label{3.3}%
  \frac{d \sigma_{qg}}{d t} &=& 2 \pi \frac{\alpha_s^2}{ s^2} (s^2 + u^2) \left[\frac{1}{t^2} -
  \frac{N_c^2-1}{2\, N_c^2 \,s u}\right],
  \eea
where $\alpha_s$ is the QCD coupling constant and $s,\, t,\, u$ are Mandelstam's variables.

As is seen, the cross sections are divergent at small  momentum transfer. So,
only leading terms of the  $1/t^2$ order are kept in \re{3.1}-\re{3.3} which
dominate for gluon/quark collisions within the Landau CI approximation.
In this approximation one can obtain the Landau-like CI:
\bea%
\label{3.9}%
C_a(\vvv{p_a},t) &=& \frac{\partial}{\partial p_{a  \alpha}}
\sum_b \int d^3p_b B_{\alpha \beta}(\vvv{p_a},\vvv{p_b})\times
\nonumber
\\ && \left[ \frac{\partial f_a}{\partial p_{a  \beta}}f_b(1 \pm f_b)
 -\frac{\partial f_b}{\partial p_{b  \alpha}}f_a(1 \pm f_a)
\right], \eea %
where $B$ is a kernel. This kernel is further simplified  for massless particles~\ci{p}:
\bea%
\label{3.24}%
B_{\alpha \beta}(\vvv{p}_a, \vvv{p}_b)&=& \xi_{a b}\left[
(1-\vvv{v_a}\vvv{v_b})\delta_{\alpha \beta}+v_{a \alpha} v_{b
\beta} +
v_{b \alpha} v_{a \beta} \right],%
\eea%
where $v_{a \alpha}=p_{a \alpha}/\omega(p_a, t)$, $\xi_{ab}=2\pi \alpha^2 G_{ab} L$.
Coefficients $G_{ab}$ are  defined by corresponding cross sections:
\bea
G_{gg} = \frac{N^2_c}{N^2_c-1}\,, \qquad G_{qg} = \frac{1}{4}\,, \qquad
G_{qq} = \frac{N^2_c-1}{8N_c^2}\,.
\eea

\section{Calculation results}

The system of three differential equations \re{2.19}-\re{2.21}, describing self-consistently the
evolution of distribution functions and fields, is closed
and can be solved numerically as the initial Cauchy problem. We choose the zero initial conditions
for distribution functions of bosons and fermions and
for a non-zero initial value of the  chromo-electric field strength $e \sigma(t=0)$.
These conditions correspond to the FTM, where colliding ions generate a large-amplitude  field
at passing ions through each other. This field is decaying due to creation of quarks and gluons.

The analysis of abundance evolution  within the pQCD model shows the dominant role of gluons~\ci{Muell}.
But as is seen from Fig.1a, it is not the case for the model considered~: Quark production  dominates over
gluons during few first $fm/c$ in the very early stage of interaction. This "fermion dominance"
 originates from spin effect in the source term \re{2.12} suppressing low $P_{\|}$ gluons~\ci{YaF01}.
 In the early stage of a collision
 when the created particle  density is still small, in the phase space there are many free states for the final state
 of created particles and the influence of the Pauli blocking  is not so essential. However,
 with subsequent density increase, the fermion creation is suppressed by the factor $(1- 2f_-)$ due to occupied states
 and bosons are enhanced by the $(1+2f_+)$ factor. Therefore, only at later time the system evolution  resembles that in
 the pQCD model exhibiting the dominant gluon production.

The mass ratio $m_+/m_-$  dependence of the fermion dominance time $\tau_f$  is of great interest.
As shown in Fig.1b, $\tau_f \approx const$  in the range of  $m_+/m_- \lsim 0.5$ corresponding
 to non-zero  fermion dominance time even in the limit of zero boson mass. However, $\tau_f \to \infty$
 when the  boson mass equals to or exceeds the  fermion mass.

Unfortunately, the  CI \re{3.9}, taking into account mainly  hard partons, is of minor
importance  and  can not result effectively in
quantum oscillation damping of the mean field. The mean field  causes a rippling  excitation
of the distribution function and makes
impossible to  achieve equilibrium in QGP in the small momentum-transfer approximation used.
Fig.2 illustrates this fact for pure gluon plasma  and for QGP.
In the first case considered the system is close to equilibrium, but taking  into account additionally
the quarks degrees of freedom, we obtain far-of-equilibrium dynamics.
The latest work~\ci{Muell} shows that this problem is not conditioned by using the Landau CI and can not
be solved by including higher terms in evaluation of corresponding cross-sections.
Due to the fermion dominance at the early stage,  quarks cause significant changes in the gluon distribution
function and  the system evolves extremely slowly towards  a quasi-equilibrium state.

\vspace{0.5cm}

\begin{figure}
\centering
\mbox{
\subfigure[]{
  \setlength{\unitlength}{0.4mm}
\begin{picture}(150,90)
 \put(0,-6){
  \epsfig{file=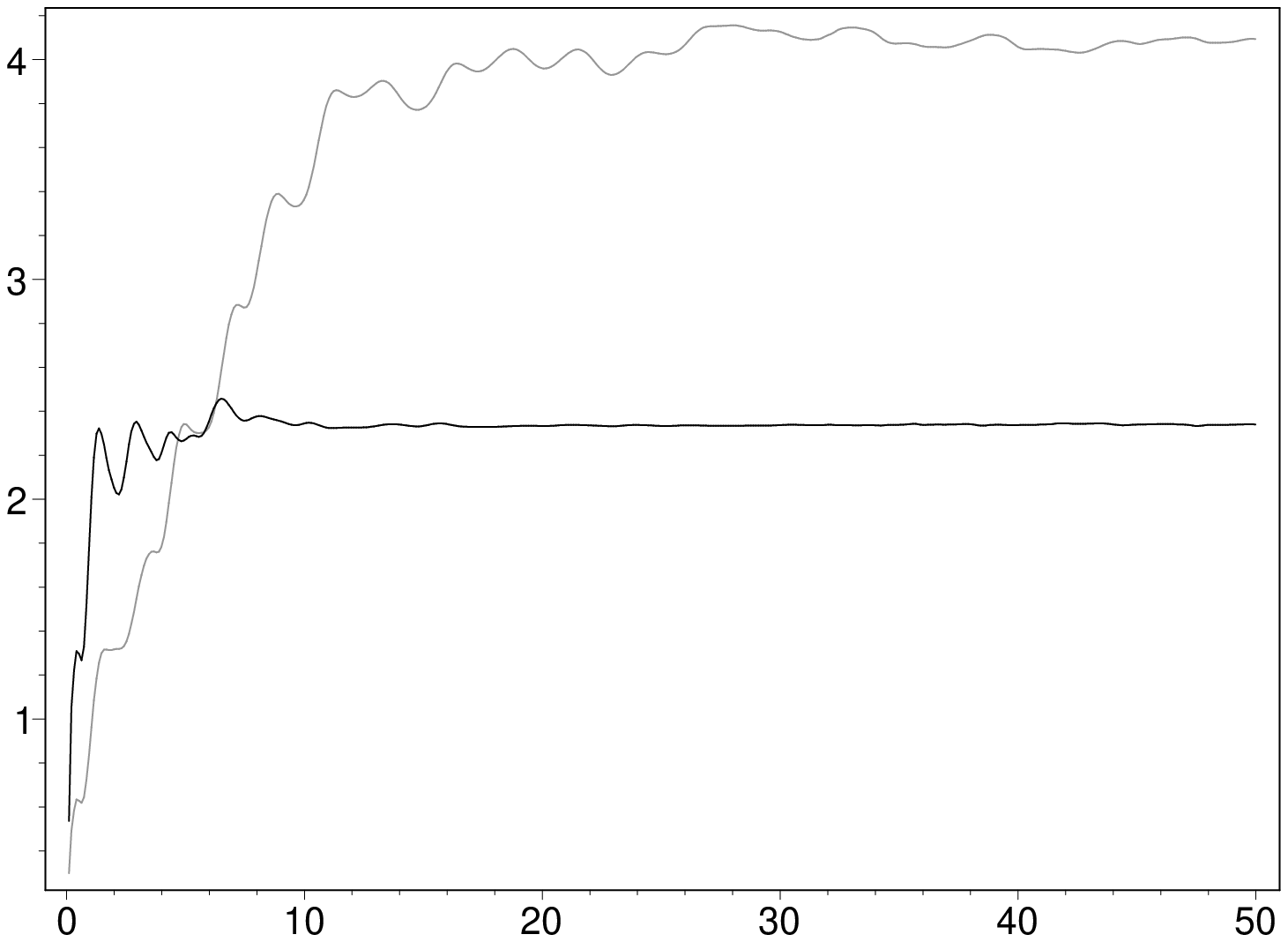, width=0.35\textwidth
   }}
 \put(1,105){{ \small $n_\pm, ~\mbox{fm}^{-3}$}}
 \put(70,-8){{ \small $ t, ~\mbox{fm/c}$}}
\end{picture}}
\hfill
\subfigure[]{
  \setlength{\unitlength}{0.4mm}
\begin{picture}(150,90)
 \put(0,0){
  \epsfig{file=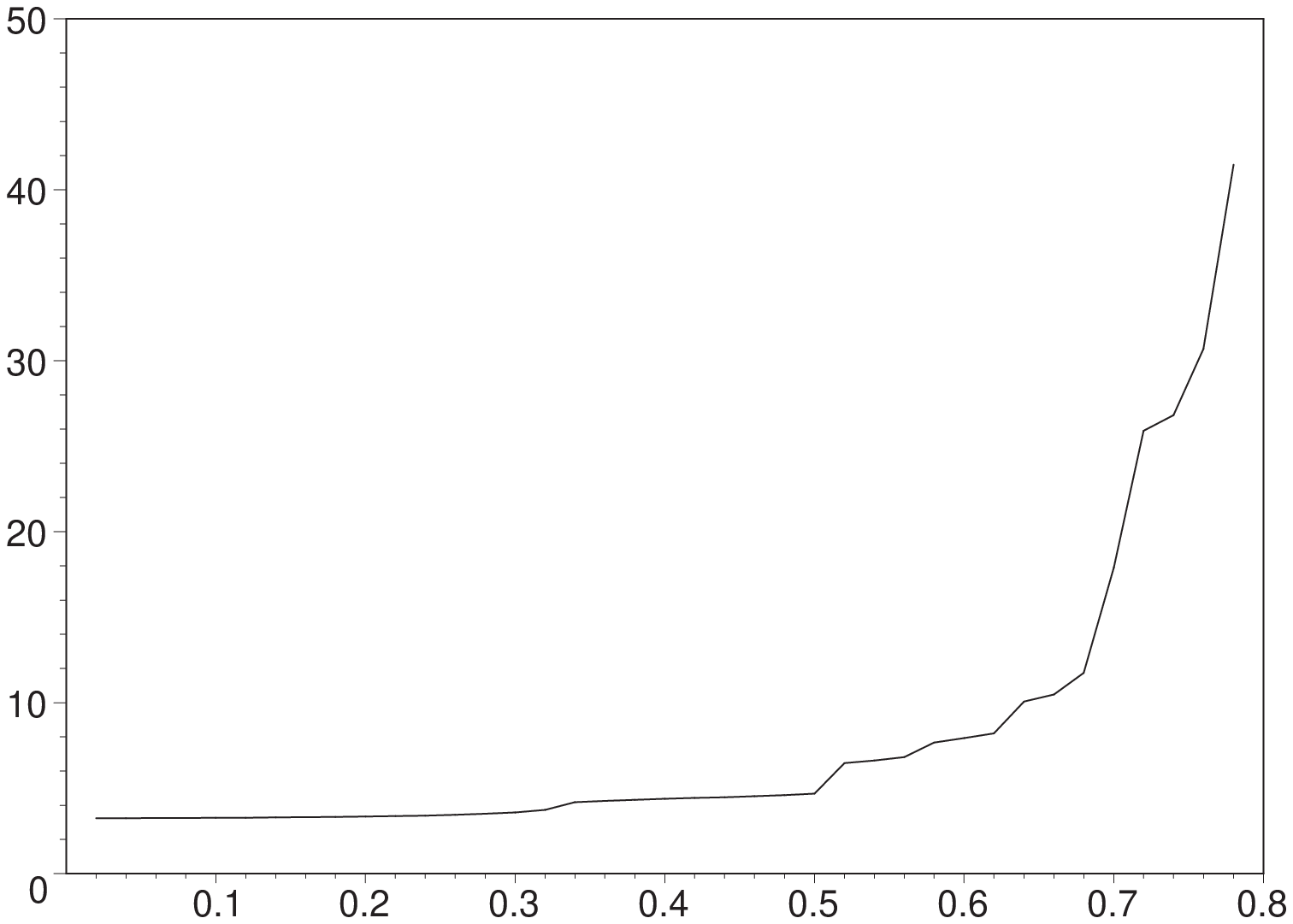, width=0.35\textwidth
   }}
 \put(-5,107){{\small $\tau_f, ~\mbox{fm/c}$}}
 \put(70,-8){{ \small ${m_+}/{m_-}$}}
\end{picture}
}}

\caption{Time dependence of number density for fermions (solid line) and for bosons (dotted line)
 (a) and the fermion dominance time $\tau_f$  as a function of the $m_+/m_-$ ratio (b). All results
 are presented for the system with the initial field strength $e \sigma(t=0)=10 \ GeV^4$.}
\end{figure}

\vspace{0.5cm}

\begin{figure}[htp]  
\centering
\mbox{
\subfigure[]{
 \setlength{\unitlength}{0.4mm}
\begin{picture}(150,90)
 \put(0,0){
  \epsfig{file=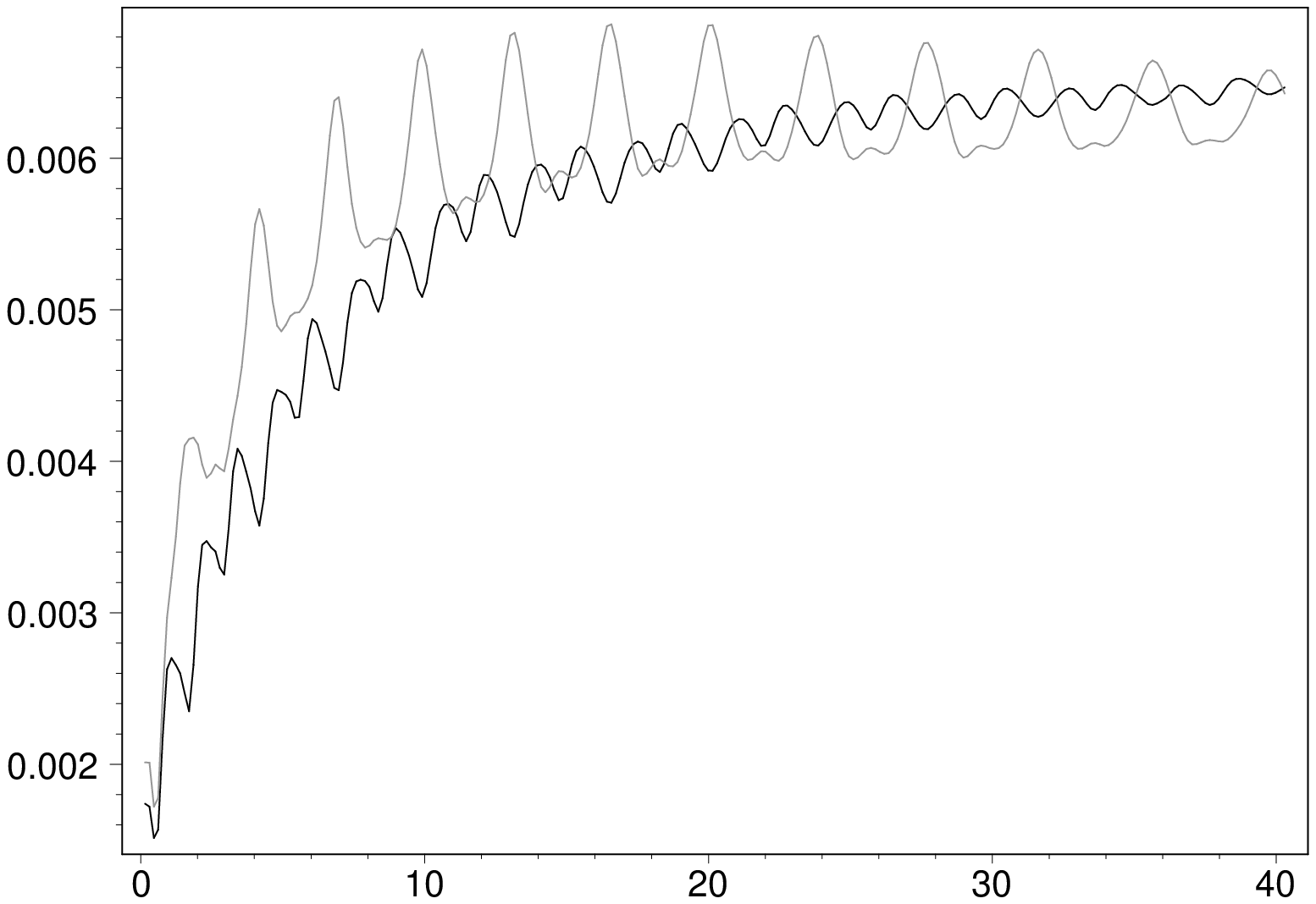, width=0.35\textwidth
   }}
 \put(1,105){{ \small $P, ~\mbox{GeV}^{4}$}}
 \put(70,-8){{ \small $ t, ~\mbox{fm/c}$}}
\end{picture}}

\subfigure[]{
  \setlength{\unitlength}{0.4mm}
\begin{picture}(150,90)
 \put(0,-2.5){
  \epsfig{file=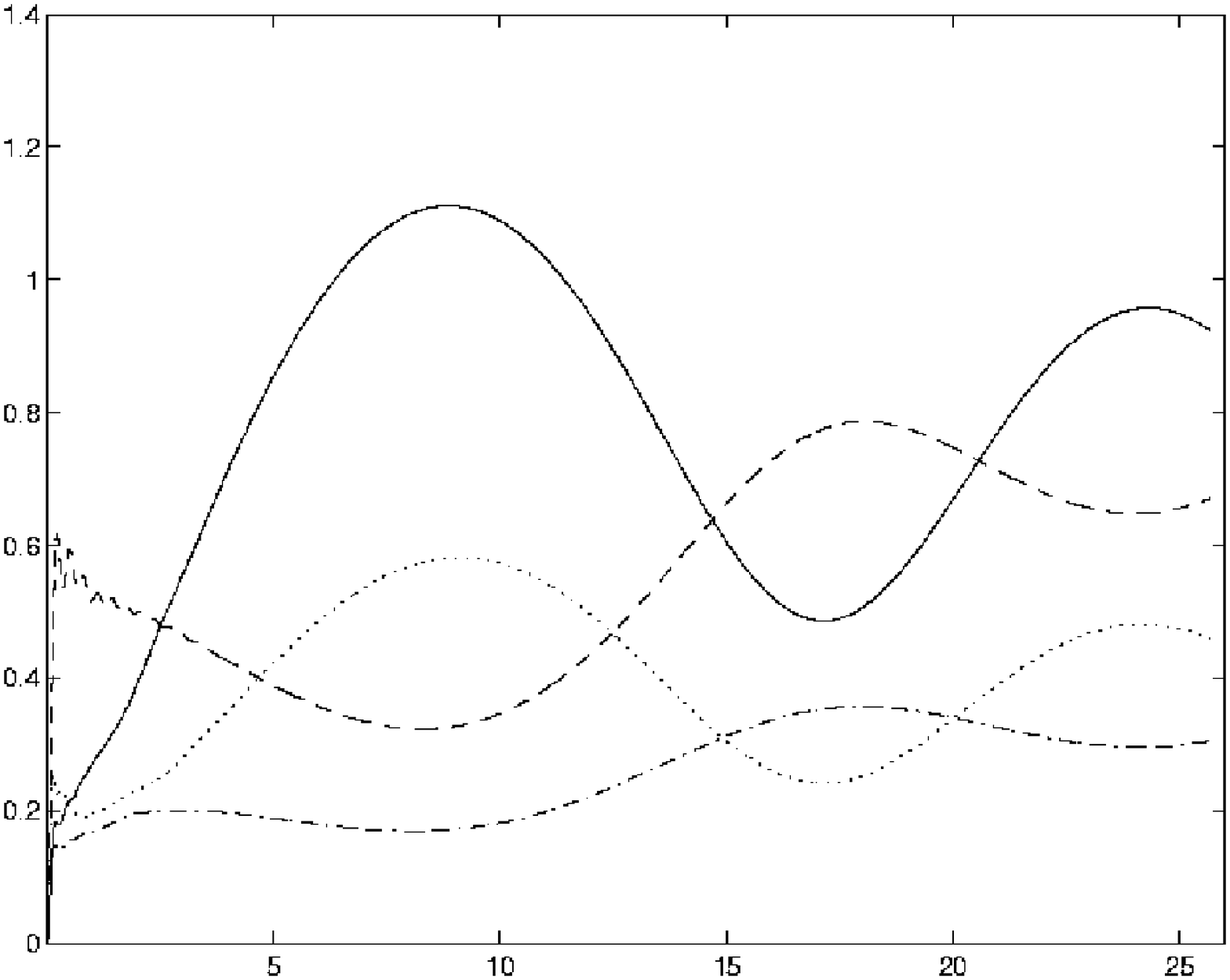, width=0.32\textwidth
   }}
 \put(-5,107){{\small $P, ~\mbox{GeV}^{4}$}}
 \put(70,-8){{ \small ${m_+}/{m_-}$}}
 \put(50,84){{\tiny ${ P^q_L}$}}
 \put(50,48){{\tiny ${P^g_L}$}}
 \put(50,32){{\tiny ${ P^q_T}$}}
 \put(50,21){{\tiny ${P^g_T}$}}
\end{picture}
}}
\caption{Time dependence of longitudinal (solid line) and transverse (dotted line) pressure for pure gluon system (a)
  and that for different parton components in quark-gluon plasma (b).}
\end{figure}
%
%

\section{Conclusions}
 The self-consistent system of equations for describing early stage of heavy ions collisions has been derived.
Being solved numerically, the set of equations does not result in a fast attainment of
a quasi-equilibrium  state for the system under discussion. To get a reasonable estimate for
the relaxation time of quark-gluon plasma one needs to increase the cross sections in few times what
effectively would correspond to accounting for radiative pertubative processes and
scatterings of nonperturbative partons, as well.
Instead of the gluons dominance predicted by pQCD models, we observe the quark dominance in the very
early stage of particle production at the time scale of few $fm/c$ corresponding to QGP evolution.
If  expansion of excited matter is included into consideration,  this time $\tau_f$ should be even longer.

One should stress that the KE method used  is rather simplified with respect to
the CI \re{3.9} in KE  \re{2.11}. This Boltzmann-like integral is obtained in the nearest order approximation
of the gradient expansion for the distribution function, neglecting all coherence effect to be related to
the mean field. As expect, the consecutive dynamical approach to the
CI problem can lead to an important correction of the result discussed~:  Strong quasiclassical fields
have to influence on quantum fluctuations of quark and gluon fields.
This problem is a real challenge to the kinetic theory of a
particle-antiparticle plasma in a strong field.

\vspace*{5mm}
Useful discussions with  A.V.~Prozorkevich and D.V.~Vinnik are acknowledged.
This work was supported in part  by the Education Ministry of Russian
 Federation under grant N E00-33-20 as well as
by DFG (project 436 RUS 113/558/0) and RFBR (grant 00-02-04012).

\end{document}